# Exploring anti-reflection modes in disordered media

Moonseok Kim[1], Wonjun Choi[1], Changhyeong Yoon[1], Guang Hoon Kim[2], Seung-hyun Kim[3],

Gi-Ra Yi[3], Q-Han Park[1] and Wonshik Choi[1,*]

[1]*Department of Physics, Korea University, Seoul 136-701, Korea*

[2]*RSS Center, Korea Electrotechnology Research Institute, Seoul, 121-835, Korea*

[3]*School of Chemical Engineering, Sungkyunkwan University, Suwon, 440-746, Korea*

**Sensing and manipulating targets hidden under scattering media are universal problems that take place in applications ranging from deep-tissue optical imaging to laser surgery[1-4]. A major issue in these applications is the shallow light penetration caused by multiple scattering that reflects most of incident light. Although advances have been made to eliminate image distortion by a scattering medium[5-7], dealing with the light reflection has remained unchallenged. Here we present a method to minimize reflected intensity by finding and coupling light into the *anti-reflection modes* of a scattering medium. In doing so, we achieved more than a factor of 3 increase in light penetration. Our method of controlling reflected waves makes it readily applicable to *in vivo* applications in which detector sensors can only be positioned at the same side of illumination and will therefore lay the foundation of advancing the working depth of many existing optical imaging and treatment technologies.**

When waves travel through disordered media such as ground glass and skin tissues, they are scattered multiple times and steadily lose momentum in the original incident direction. Large fraction of the incoming energy bounces back as a consequence, which leads to the attenuation of transmission. Considering that waves have to penetrate through scattering layers for the efficient sensing and manipulation of the targets embedded in them, this attenuation in transmission has been a limiting factor for the working depth of various optical techniques. For example, poor light penetration through skin tissues is a major barrier to the improvement of treatment depth in phototherapy [1,8,9] and to the imaging depth of various imaging techniques including photoacoustic tomography [2], optical projection tomography [3] and diffuse optical tomography [4,10]. Similarly, optical scattering by the skull and brain tissues hinder light activation of neural activity in optogenetics [11-13].

In recent studies, much effort has been devoted to controlling wave propagation through a disordered medium using wavefront shaping and sensing techniques [5,6,14-26]. In particular, advances have been made for enhancing light transmission through scattering layers [16,21]. The main strategy of these previous studies has been to find the specific shape of incident wave that induces strong constructive interference at the far side of the disordered medium. This approach can be so effective that even the complete elimination of the attenuation is possible in theory [27-29]. However, all such previous efforts have required detectors positioned on the opposite side of a scattering medium to the incident wave for the direct monitoring of the transmitted wave. This makes them unsuitable for *in vivo* applications in which detectors can only be positioned on the same side as the incident wave and the monitoring of reflected wave is the only possible mode of detection.

In this study, we propose a method of interrogating reflected waves so as to find specific incident wave patterns that induce the destructive interference of reflected waves. We demonstrated that shaping of an incident wave into these what we call *anti-reflection modes* indeed reduces reflectance, and that, due to the energy conservation, the reduced reflection energy is transferred to the opposite side of the disordered media. We observed more than a factor of 3 increase in light transmission after the coupling of the incident waves to the anti-reflection modes. The working principle of this approach is similar to that of the anti-reflection coatings covering the eyeglasses and display screens in that destructive wave interference is used for minimizing unwanted reflections. However, we achieved the same effect as anti-reflection coating by simply choosing a proper pattern of incident wave rather than physically attaching a thin film. This approach, which delivers light deeper into the scattering media, will be particularly useful for various *in vivo* applications where reflection mode of detection is the only option.

In order to identify the anti-reflection modes of a disordered medium, a highly detailed measurement of the medium's response to the incident wave is required. The best possible way is to record reflected waves for all possible input waves. For this purpose, we constructed an experimental setup (See Fig. 1a for schematic setup and Fig. S1 for detailed setup) in which a clean planar wave from a He-Ne laser (633 nm in wavelength) was sent to the medium at various incident angles covering two orthogonal polarizations. For each incident angle with a certain polarization state ($\theta_x$, $\theta_y$, $p$), where ($\theta_x$, $\theta_y$) indicates illumination angle and $p$ stands for either horizontal (H) or vertical (V) polarization state, we took images of the H and V components of the reflected waves. To obtain a complex field map of the reflected wave containing both amplitude and phase, we used an interferometric microscope [6]. Therefore, the acquired image is a complex field map, $E(x, y, p; \theta_x, \theta_y, p)$, where ($x$, $y$) is a position vector in the sample plane. The illumination area was set to be $11\times11$ $\mu m^2$ and the angular range covered was up to a numerical aperture of 0.8. The minimum required number of incident angles for the complete coverage of the given area and numerical aperture is 640. Here we used 800 angles of incident wave, which is a slight oversampling, for enhancing the signal to noise in the matrix construction. The collection area ($32\times32$ $\mu m^2$) for the reflected wave was set larger than the illumination area to capture waves that laterally diffuse away from the illumination area. The angular coverage of collection was set close to the numerical aperture of 1.0 to capture all the angularly scattered waves. As two polarization states were covered for both input and output waves, in total 3,200 images of reflected waves were recorded. As a disordered medium, we used randomly distributed $TiO_2$ particles on a coverslip, whose average transmittance ranges between 10% and 15 %. The thickness of the medium is about $8 \pm 2$ μm and the transport mean free path is $0.5 \pm 0.2$ μm such that the transmitted light is scattered hundreds of times on average.

We then processed the 3,200 images to construct a so-called reflection matrix $R$. The input and output bases of this matrix are $n = (\theta_x, \theta_y, p)$ and $m = (x, y, p)$, respectively. The matrix element $r_{mn}$ connects the $n^{th}$ coordinate in the input to the complex amplitude (both amplitude and phase) of the wave at the $m^{th}$ coordinate in the output. Thus the reflected wave $E_m^{ref}$ at the $m^{th}$ coordinate is given by $E_m^{ref} = \sum_n r_{mn} E_n^{in}$, where $E_n^{in}$ is the complex amplitude of the incident wave at the $n^{th}$ input coordinate. Below is the procedure of the matrix construction. For each input coordinate, $(\theta_x, \theta_y, p)$, there are two sets of reflection images, $E(x, y, H; \theta_x, \theta_y, p)$ and $E(x, y, V; \theta_x, \theta_y, p)$. We converted each image into a vector by appending each column in the reflected image to the end of the preceding column. We then appended the image vector for V polarization to the one for H polarization. The resulting merged reflection image vector was assigned to a column of the reflection matrix. By repeating the same procedure for all input coordinates, the reflection matrix $R$ was constructed. The figures 2a and 2b show the amplitude and phase parts of the constructed reflection matrix. The column index is the input coordinate covering both H polarization and V polarization, and the row index indicates output coordinate.

From the constructed reflection matrix, we extracted anti-reflection modes of a disordered medium. According to random matrix theory (RMT), the singular value decomposition of a reflection matrix allows the discovery of the eigenchannels [27]. Specifically, the reflection matrix was factorized into $R = U\Sigma V^*$, where $\Sigma$ is a rectangular diagonal matrix with nonnegative real numbers on the diagonal called singular values, and $V$ and $U$ are the unitary matrices whose columns are the reflection eigenchannels for the input and output, respectively. The $V^*$ denotes the conjugate transpose of the matrix $V$. The singular values in $\Sigma$ were sorted in the descending order. We assigned eigenchannel index from 1 to $N$ for the sorted singular values, where $N = 1280$ is the total number of eigenchannels. Therefore, the $i^{th}$ column in $V$ is the $i^{th}$ reflection eigenchannel on the input plane, and the individual element in the column is the complex amplitude of the corresponding $(\theta_x, \theta_y, p)$. The square of the singular value is called reflection eigenvalue, which is the expected reflectance of the corresponding eigenchannel. Green square dots in Fig. 4a show the eigenvalue distribution calculated from the measured reflection matrix where we observed monotonic decrease in the eigenvalue. We define the eigenchannel with smallest eigenvalue, whose eigenchannel index corresponds to 1280, as anti-reflection modes of the medium.

We physically shaped an incident wave into a reflection eigenchannel, i.e. a column of $V$, using the setup shown in Fig. 1b (See Fig. S1 for detailed setup). Two spatial light modulators (SLM I and SLM II) that operate at phase-only control mode were used to shape the H and V polarization components of individual eigenchannel. They were carefully aligned, both laterally and axially, to be overlapped at the input plane of the sample with a positioning accuracy of 150 nm. Figures 3a and 3b show the intensity maps of the incident waves for the first eigenchannel and the anti-reflection mode, respectively. We then measured the reflected wave images for each case (Figs. 3d and 3e). As can be seen, the total reflection intensity of the first eigenchannel is much higher than that of the anti-reflection mode, and the reflection intensity of the anti-reflection mode, the least reflecting mode of the medium, was highly attenuated. We measured the intensity map of the transmitted waves for each reflection eigenchannel by positioning a camera on the opposite side of the medium (Figs. 3g and 3h) and observed that the total transmission intensity of the anti-reflection mode is significantly higher than that of the first eigenchannel. As a point of reference, we sent a planar wave of normal incidence to the medium (Fig. 3c) and recorded the intensity maps of the reflected (Fig. 3f) and transmitted (Fig. 3i) waves. Compared with the transmittance of this normally incident plane wave, transmittance of the anti-reflection mode was enhanced by a factor of 3. This result proves that we have enhanced wave penetration through a highly scattering medium with the measurement of reflected waves, not the transmitted waves, which makes our approach realistic for many practical applications.

We recorded the reflectance for many of the representative eigenchannels (Fig. S2). The reflectance monotonically decreased as the eigenchannel index was increased (blue circular dots in Fig. 4a). The measured reflectance for the anti-reflection mode was 35.7 %, which is far smaller than the reflectance of uncontrolled input wave of 71.1 %. As a consequence of the net decrease of reflectance by 35.6 %, the transmission was increased from an average transmittance of 12.8 % to 38.8 %. We found that not all of the reduced reflection energy was converted to the transmission energy. In fact, there was an approximate 9.6 % loss due to the use of slab geometry, rather than the ideal waveguide geometry, which allowed some of the scattered waves to leak out of collection range. However, this limitation is consistent with real applications such as the interrogation of targets under biological skin tissues because the scattering layers are in slab geometry by nature. According to our numerical analysis based on RMT (Fig. S4), the expected enhancement factor was around 4 after accounting for the effect of this leakage, which is close to the experimentally achieved enhancement factor of 3. In Fig. 4a, the degree of decreased reflectance is somewhat lower than that predicted by the reflection eigenvalues. This difference is largely due to the phase-only control of SLMs and the system's mechanical drift.

The increase in transmission through the coupling of incident wave into the anti-reflection mode implies that the incident wave has preferentially coupled to those transmission eigenchannels with large transmittance. We explored the explicit connection between reflection eigenchannels and their transmission counterparts. We recorded a transmission matrix for the same medium (Fig. S3) and obtained the transmission eigenchannels. For the reflection eigenchannel of minimum eigenvalue, i.e. the anti-reflection mode of the medium, we obtained cross-correlation with various transmission eigenchannels (red circular dots in Fig. 4c). Here, we found that the transmission eigenchannels with larger transmission eigenvalues (i.e. of smaller transmission eigenchannel indices) preferentially contribute to the anti-reflection mode. On the other hand, in the case of the reflection eigenchannel of large eigenvalue, the cross-correlation shows an increase as the eigenchannel index of transmission increases. This clearly shows that the coupling of incident waves to the anti-reflection mode of the medium is equivalent to finding transmission eigenchannels of high transmittance.

In conclusion, we demonstrate the enhancement of wave penetration to the scattering medium by coupling light into anti-reflection modes of the medium. For the sensing, imaging and manipulation of targets obscured by highly scattering layers, it is essential to send interrogating waves with sufficient intensity to the target sites. Therefore, our study can potentially advance many practical applications for which reflection loss caused by the scattering layers is a limiting factor. For example, our method can enhance the working depth of photoacoustic imaging, diffuse optical tomography, and *in vivo* glucose sensing [30]. Similarly, by administering an intense light source and minimizing loss in this way, the treatment depth of various phototherapies can be increased, and the optical manipulation of cells embedded in the tissues can be made efficient.

**Methods**

In order to record both reflection and transmission matrices of a disordered medium, we constructed an interferometric microscope working in both reflection and transmission modes. Two spatial light modulators were installed in the illumination beam path to control two orthogonal polarization states. A wave plate was inserted in the reference beam path to set the polarization state of reference beam to be either horizontal or vertical. The polarization state of the reference beam determines the polarization component of detected waves. In finding the eigenchannels of the medium, we performed singular value decomposition using Matlab built-in code. As a disordered medium, we used $TiO_2$ particle (Sigma Aldrich) layers. In order to prepare relatively uniform layers, we made a solution of $TiO_2$ in ethanol and spread the layer on the slide glass (or cover glass) by using air spray. We also made a similar sample with better uniformity in thickness using a spin coating method.

**Acknowledgements**

This research was supported by the Basic Science Research Program through the National Research Foundation of Korea (NRF) funded by the Ministry of Education, Science and Technology (2011-0016568), the National R&D Program for Cancer Control, the Ministry of Health & Welfare, South Korea (1120290), the Seoul metropolitan government, Korea under contract of R&BD Program WR100001, and Korea MOTIE grant (NO. Sunjin-2010-002).

**Supplementary Information**

Supplementary text and Figures S1-S4.

**Figure legends**

**Figure 1. Experimental schematic for the recording of a reflection matrix and implementation of reflection eigenchannels.** The diagrams shown in (**a**) and (**b**) are conceptual (See Fig. S1 for the detailed experimental setup) **a**, A planar wave with a wavelength of 633 nm was sent to a disordered medium (S) at an illumination angle of ($\theta_x$, $\theta_y$) with either horizontal (H; red) or vertical (V; sky-blue) polarization. The reflected wave was divided into two polarization states by a setup equivalent to positioning a polarizing beam splitter (PBS), and the transmitted (reflected) wave at the PBS having H (V) polarization was recorded by a camera I (II). **b**, A 45$^\text{o}$ linearly polarized plane wave having both H and V polarization components aimed at a PBS. The vertically polarized wave reflected at the PBS illuminated a spatial light modulator, SLM I, which rotated the polarization state by 90 degrees and shaped the H polarization component of a reflection eigenchannel. Then the wave reflected by the SLM I was guided to the disordered medium. Simultaneously, the SLM II shapes V polarization component of the same reflection eigenchannel.

**Figure 2. Experimentally recorded reflection matrix of a disordered medium. a**, Amplitude part of the reflection matrix. Column indices correspond to input coordinates, and were assigned with an ascending order of the azimuthal angle for H polarization and then the same order of angle for V polarization. Row indices indicate spatial coordinate (x,y) for both polarization states of reflected waves. Color bar indicates amplitude in an arbitrary unit. **b**, The phase part of the reflection matrix. Color bar indicates phase in radians. Yellow and black dashed lines were drawn for visual guidance.

**Figure 3. Shaping incident wave into reflection eigenchannels. a**, **b** and **c**, Intensity maps of experimentally shaped incident waves for the first eigenchannel, 1280$^{th}$ eigenchannel and normally incident plane wave, respectively. The H and V polarization components were superposed at the incident plane. **d**, **e** and **f**, Experimentally recorded intensity maps of reflected waves for the cases of (**a**), (**b**) and (**c**), respectively. **g**, **h** and **i**, Measured intensity maps of transmitted waves for the case of (**a**), (**b**) and (**c**), respectively. Scale bar, 5 µm. The color bar next to (**c**) indicates intensity in an arbitrary unit and applies also to (**a**) and (**b**). The other two color bars follow the same rule.

**Figure 4. Reflectance and transmittance of various reflection eigenchannels. a**, Reflectance of experimentally implemented reflection eigenchannels (blue circular dots) and those predicted from the eigenvalues of the measured reflection matrix (green square dots). Black dots indicate mean reflectance of the medium. **b**, Transmittance of the same reflection eigenchannels (red circular dots) implemented in (**a**). Black dots indicate mean transmittance of the medium. **c**, Relation between reflection and transmission eigenchannels**.** Reflection and transmission matrices were measured for the same disordered medium, and their respective eigenchannels were obtained by performing the singular value decomposition. Red circular dots: absolute square of normalized cross-correlation between the first reflection eigenchannel and all the transmission eigenchannels. The absolute squares for every 30 transmission eigenchannels were added. Blue square dots: the same as the red circular dots, but for the 1280$^{th}$ reflection eigenchannel.

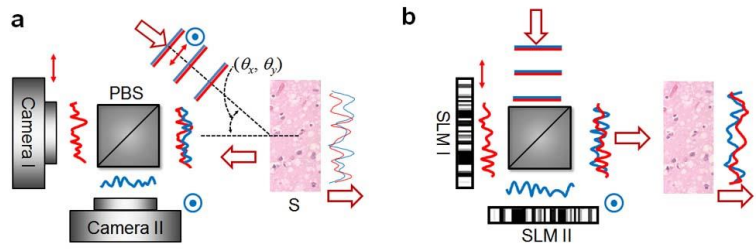
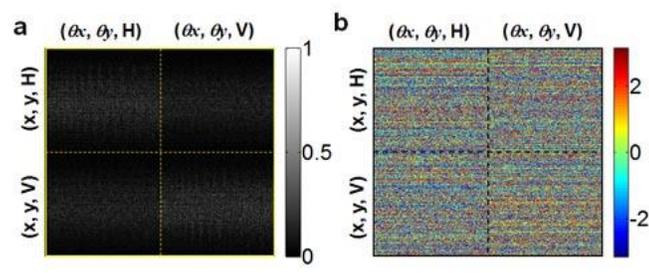
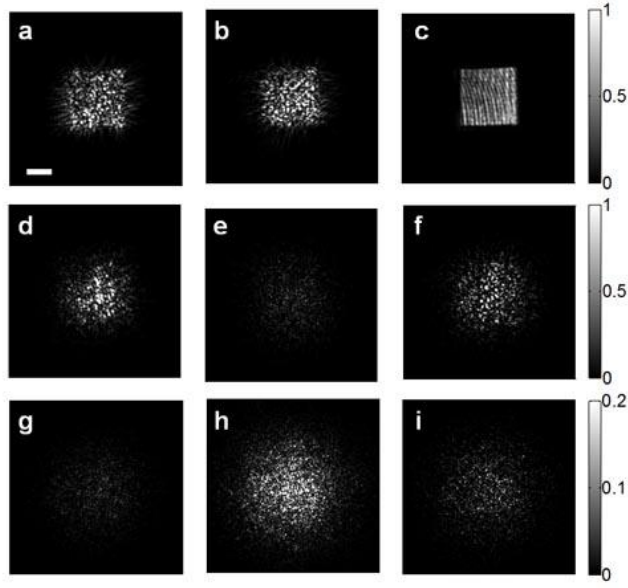
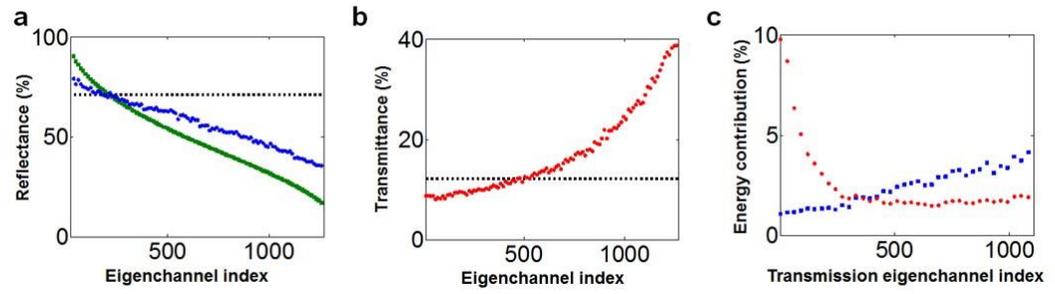